\documentstyle{mn}
\begin{document}

\def\simlt{\mathrel{\rlap{\lower 3pt\hbox{$\sim$}}
        \raise 2.0pt\hbox{$<$}}}

\def\simgt{\mathrel{\rlap{\lower 3pt\hbox{$\sim$}}
        \raise 2.0pt\hbox{$>$}}}

\title[Clustering Properties and Gravitational
Lensing of Forming Spheroidal Galaxies]
{Clustering Properties and Gravitational Lensing of Forming
Spheroidal Galaxies}

\author[F.~Perrotta et al.]
{\parbox[t]{\textwidth}{F.~Perrotta$^{1,2}$, M.~Magliocchetti$^{2}$,
C.~Baccigalupi$^{2}$, M.~Bartelmann$^{3}$, G.~De Zotti$^{1}$,
G.L.~Granato$^{1}$, L.~Silva$^{4}$, L.~Danese$^{2}$}\vspace*{6pt} \\
$^{1}$ Osservatorio Astronomico di Padova, Vicolo
dell'Osservatorio 5, I-35122 Padova, Italy\\
$^{2}$ SISSA/ISAS, Via Beirut 4, I-34014 Trieste, Italy\\
$^{3}$ Max-Planck Institut f\"ur Astrophysik, P.O.~Box 1317, D--85741 Garching, Germany\\
$^{4}$ Osservatorio Astronomico di Trieste, Via G. B. Tiepolo, 11,
34131 Trieste, Italy}
\maketitle

\begin{abstract}

We show that the predictions of the recent physical model by
Granato et al. (2001) are fully consistent with the available
estimates of the clustering properties of galaxies at (sub)-mm
wavelengths. The model is then exploited to compute the expected
power spectrum of fluctuations due to clustering at the
frequencies of the High Frequency Instrument (HFI) of ESA's {\sc
Planck} satellite. The clustering signal is found to be detectable
in regions of low interstellar dust emission. A further
distinctive prediction of the adopted model is a remarkably high
fraction of gravitationally lensed sources at bright mm/sub-mm
fluxes. In fact, according to the adopted model, only thanks to
gravitational lensing, a significant number of high-$z$ forming
spheroidal galaxies will be detectable by large area, shallow
surveys at mm/sub-mm wavelengths, such as those carried out by
P{\sc lanck}/HFI. Allowing for other source populations, we find
that the fraction of gravitationally lensed sources at bright
mm/sub-mm fluxes is expected to be up to $\simeq 40\%$.

\end{abstract}

\begin{keywords}gravitational lensing -- galaxies: formation --
large scale structure of the Universe -- submillimetre: galaxies

\end{keywords}

\section{Introduction}

\label{intro}

The formation and early evolution of massive spheroidal galaxies
is still a most controversial issue. In the general framework of
the hierarchical clustering picture for structure formation in
Cold Dark Matter (CDM) cosmologies, two scenarios are confronting
each other. According to the ``monolithic'' scenario, a large
fraction of giant elliptical galaxies formed most of their stars
in a single gigantic star-burst at high redshift, and then
essentially underwent passive evolution. This approach is in
contrast with the ``merging'' scenario, backed up by semi-analytic
galaxy formation models (e.g. Kauffmann 1996; Baugh et al. 1998),
wherein large galaxies mostly formed via mergers of smaller
galaxies that, at the time of merging, had already converted part
of their gas into stars. These two scenarios yield very different
predictions of the evolution of large ellipticals: if
``monolithic'' collapse is assumed, the comoving number density of
such galaxies remains essentially constant with redshift, while
their bolometric luminosity increases with look-back time out to
their epoch of formation. On the contrary, according to the
``merging'' scenario, the comoving density of large ellipticals
decreases with increasing redshift.

The evolution of the comoving density in the optical band has
however proven difficult to measure, and discordant results have
been reported. A deficit of old ellipticals at $z\simgt 1$, when
compared to predictions from pure luminosity evolution models, has
been claimed by several groups (Kauffmann et al. 1996; Zepf 1997;
Franceschini et al. 1998; Barger et al. 1999b; Menanteau et al.
1999; Treu \& Stiavelli 1999; Rodighiero et al. 2001), while
others found evidence for a constant comoving density up to $z\sim
1.5$ (Totani \& Yoshii 1997; Benitez al. 1999; Broadhurst \&
Bowens 2000; Scodeggio \& Silva 2000; Cohen 2001).

Furthermore, Daddi et al. (2000) find that the observed surface
density of Extremely Red Objects (EROs)-- the bulk of which appears
to be passively evolving ellipticals -- is consistent with
predictions from the ``monolithic'' scenario.

The results of SCUBA and MAMBO surveys at sub-mm/mm wavelengths
severely challenge the ``merging'' scenario. The most exhaustive
investigation carried out so far (Granato et al. 2000) combines
the semi-analytic galaxy formation model by the Durham group (Cole
et al. 2000) with the state-of-the-art spectro-photometric code by
Silva et al. (1998) which includes reprocessing of starlight by
dust. It turns out that, while this approach successfully accounts
for a large variety of observables (spectral energy distributions
of galaxies of different morphological classes, luminosity
functions over a wide wavelength range, from UV to far-IR, and
more), it nevertheless falls short by a substantial factor when
trying to account for the mm/sub-mm counts. Similarly, Devriendt
\& Guiderdoni (2000), who used a totally independent semi-analytic
model, were forced to introduce an ad-hoc galaxy population to
reproduce such counts.

The main problem stems from the fact that, although only a handful
of redshifts of SCUBA/MAMBO sources have been reliably measured so
far, there is growing circumstantial evidence that many (perhaps
most) of them are ultraluminous star-forming galaxies at $z \simgt
2$ (Smail et al. 2000; Dunlop 2000; Carilli et al. 2000; Bertoldi
et al. 2000; Scott et al. 2001; Fox et al. 2001). Furthermore, the
inferred star-formation rates are very high (typically from a few
hundreds to $\sim 1000\,\hbox{M}_\odot\,\hbox{yr}^{-1}$),
consistent with those required to build the stellar populations of
massive ellipticals on a timescale $\simlt 1\,$Gyr.

A detailed, astrophysically grounded, model which fully accounts
for both the SCUBA/MAMBO counts in the framework of
hierarchical clustering scenarios, and for the main aspects of the
chemical evolution of spheroidal galaxies (stellar metalicity,
luminosity-metalicity relationship, $\alpha$ enhancement) has
been recently presented by Granato et al. (2001). Key ingredients
of this model are: i) a tight relationship between formation and
early evolution of spheroidal galaxies and active nuclei hosted
at their centres (note that first evidences for a SCUBA/AGN connection
have recently been found by Almaini et al. 2001);
ii) once the effects of cooling
and heating (the latter being mostly due to stellar
feedback) are properly taken into account, the timescale for star
formation within virialised dark matter (DM) halos turns out to be
relatively short for massive spheroidal galaxies while it is longer in the case
of less massive halos, where the feedback from supernovae (and/or
the active nucleus) slows down the star formation and can also
expel a significant fraction of the gas. It therefore follows that the
canonical
scheme implied by the hierarchical CDM scenario -- small clumps
collapse first -- is reversed in the case of baryons.

During the active star-formation phase, large spheroidal galaxies
show up as luminous sub-mm sources, accounting for the SCUBA and
MAMBO counts. The subsequent evolution is essentially passive,
except possibly for minor mergers, much like that expected in the
``monolithic'' scenario.

As mentioned by Granato et al. (2001), the model implies that
SCUBA/MAMBO galaxies are highly biased tracers of the matter
distribution. The model also establishes well defined links
between these galaxies and other galaxy populations such as Lyman
Break Galaxies (LBGs) and Extremely Red Objects (EROs).  Studies
of their clustering properties therefore provide a test for it. A
first discussion of this aspect was offered by Magliocchetti et
al. (2001a, henceforth MA2001). In this paper we extend the
analysis and compare the clustering properties predicted by the
model with recent observational measurements at $850\,\mu$m and
$170\,\mu$m.

Another distinctive feature of this model is the extremely steep,
essentially exponential decline of the bright tail of mm/sub-mm
counts of forming spheroids. This is due to the fact that, according to the
adopted model, the bulk of star formation in such objects is essentially
complete at $z \simeq 2$. The bright counts therefore essentially
reflect the high luminosity tail of the luminosity function. This in turn
relies on the fact that within the
hierarchical clustering scenario massive halos are exponentially
rare at high redshifts. Hints of such a steep decline can possibly
be discerned in the recent MAMBO (Carilli et al. 2000; Bertoldi et
al. 2000) and SCUBA bright surveys (Scott et al. 2001; Borys et al. 2001).

As extensively quantified and discussed in Sections 4 and 5, such
a steep decrease in the number counts for fluxes $10{\rm mJy}
\;\simlt S_{\rm 850\mu m}\simlt 100\;{\rm mJy}$ implies a large
fraction of strongly lensed galaxies to appear at bright mm/sub-mm
fluxes. This fraction is found to be larger than those worked out
for other (phenomenological) models that also successfully account
for SCUBA/MAMBO counts (Rowan-Robinson 2001; Takeuchi et al. 2001;
Pearson et al. 2001; Blain et al. 1999b).

The layout of this paper is as follows. In \S~2 we compare the
predictions of the model by Granato et al. (2001) with recent
measurements of clustering properties of SCUBA galaxies (Peacock
et al. 2000; Scott et al. 2001) and draw some conclusions on the cell-to-cell
fluctuations of the projected density of sources, found to be strong enough
to account for the discrepancies in the number counts observed in
different regions of the sky by different groups.
In \S~3 we present the power spectra of
temperature fluctuations due to clustering of forming spheroidal
galaxies in {\sc Planck}/HFI channels and derive predictions for
the power spectrum of extra-galactic background fluctuations at $170\,\mu$m,
to be compared with the signal detected by Lagache \& Puget (2000).
\S~4 gives the essential
details of our calculations on the effect of lensing on source
counts, while in \S~5 we report our predictions for counts of
lensed galaxies at mm/sub-mm wavelengths, with reference to those
covered by the {\sc Planck}/HFI instrument. Conclusions are
presented in \S~6.

Throughout the paper we will assume a Cold Dark Matter flat
cosmology with $\Omega_{\Lambda}=0.7$, $\Omega_{b}h^{2}=0.025$,
$H_0= 100h_0\,\hbox{km}\,\hbox{s}^{-1}\,\hbox{Mpc}^{-1}$ with
$h_0=0.7$ and a COBE-normalised $\sigma_8=1$. We note however that
our conclusions are only weakly dependent on the underlying
cosmological model.

\section{Clustering of high-z forming spheroidal galaxies: model predictions vs.~observations}

A crucial test for a galaxy formation model is its ability to
correctly describe the large-scale properties of the populations
under investigation. In MA2001 we have presented predictions for
the clustering of SCUBA sources and their contribution to the
total background fluctuations in the sub-mm band, based on the
model by Granato et al. (2001). The model has been proven to
provide a good fit to the clustering properties of LBGs
(Giavalisco et al. 1998), corresponding in our framework to the
early evolutionary phases of low-to-intermediate mass spheroidal
galaxies.

In this Section we will extend the analysis of MA2001 and compare
the results to the best up-to-date clustering measurements at
850~$\mu$m. We start off with the theoretical expression for the
angular two-point correlation function $w(\theta)$:

\begin{eqnarray}
w(\theta)=2\:\frac{\int\int_0^{\infty}N^2(z)\;b_{\rm eff}^2(M_{\rm
min},z)\;(dz/dx)\;\xi(r,z)\;dz\;du}{\left[\int_0^{\infty}N(z)\;dz\right]^2},
\label{eq:limber}
\end {eqnarray}
(Peebles 1980), where $\xi(r,z)$  is the spatial mass-mass
correlation function (obtained as in Peacock \& Dodds 1996; see
also Moscardini et al., 1998 and Magliocchetti et al. 2000),
$x$ is the comoving radial
coordinate, $r=(u^2+x^2\theta^2)^{1/2}$ (for a flat universe and
in the small angle approximation), and $N(z)$ is the number of
objects within the shell ($z,z+dz$). The relevant properties of
SCUBA galaxies are included in the redshift distribution of
sources $N(z)$, and in the bias factor $b_{\rm eff}(M_{\rm
min},z)$.

The effective bias factor $b_{\rm eff}(M_{\rm min}, z)$ of all the
dark matter haloes with masses greater than some threshold $M_{\rm
min}$ is then obtained by integrating the quantity $b(M,z)$ (whose
expression has been taken from Jing 1998) -- representing the bias
of individual haloes of mass $M$ -- weighted by the mass function
$n_{\rm SCUBA}(M,z)$ of SCUBA sources:

\begin{eqnarray}
b_{\rm eff}(M_{\rm min},z)= \frac{\int_{M_{\rm min}}^{\infty}
dM\;b(M,z)\;n_{\rm SCUBA}(M,z)} {\int_{M_{\rm min}}^{\infty}
dM\;n_{\rm SCUBA}(M,z)}, \label{eq:bias}
\end{eqnarray}
with $n_{SCUBA}(M,z)=n(M,z)\;T_B/t_h$, where $n(M,z)\; dM$ is the
mass function of haloes (Press \& Schechter, 1974; Sheth \& Tormen
1999), $T_B$ is the duration of the star-formation burst and $t_h$
is the life-time of the haloes in which these objects reside.

The bias factor $b_{\rm eff}$ in Eq.~(\ref{eq:bias}) and the
angular correlation function of Eq.~(\ref{eq:limber}) have then
been evaluated for three mass ranges: i) masses in the range
$M_{\rm sph}\simeq 10^9-10^{10}M_\odot$, duration of the star
formation burst $T_B\sim 2$~Gyr, and typical $850\,\mu$m fluxes
$S\simlt 1$~mJy; ii) masses in the range $M_{\rm sph}\simeq
10^{10}$--$10^{11}M_\odot$ and $T_B\sim 1$~Gyr; iii) masses in the
range $M_{\rm sph}\simgt 10^{11}M_\odot$, $T_B\sim 0.5$~Gyr,
dominating the counts at $850\,\mu$m fluxes $S\simgt 5-10$~mJy
(note that by $M_{\rm sph}$ we denote the mass in stars at
the present time).  

Furthermore, we considered two extreme values for the ratio
between mass in stars and mass of the host dark halo -- $M_{\rm
halo}/M_{\rm sph}=100$ and $M_{\rm halo}/M_{\rm sph}=10$ -- which
encompass the range covered by recent estimates for massive
objects (McKay et al. 2001; Marinoni \& Hudson 2001). Granato et al. 
(2001) found that this ratio is $20-30$ for the massive objects. 

According to our model, sources are expected to be strongly
clustered, with a clustering signal which increases for  brighter
sources and higher values of $M_{\rm halo}/M_{\rm sph}$. From the
above discussion it follows that measurements of $w(\theta)$ are
in principle able to discriminate amongst different models for
SCUBA galaxies and in particular to determine both their
star-formation rate, via the amount of baryonic mass actively
partaking of the process of star formation, and the duration of the
star-formation burst.

\begin{figure}
\vspace{7cm}  
\includegraphics{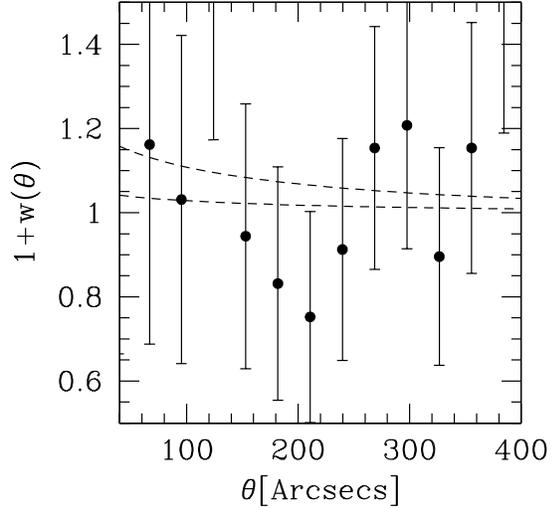}
\caption{Angular correlation function of $S\ge 5$~mJy SCUBA
sources. Model predictions are shown by the dashed curves (where
the higher one is for $M_{\rm halo}/M_{\rm sph}=100$ and the lower
one for $M_{\rm halo}/M_{\rm sph}=10$), while data points
represent the Scott et al. (2001) measurements. \label{fig:data}}
\end{figure}

Tentative evidence for positive clustering of SCUBA galaxies on scales of
$1'$--$2'$ has been found by Scott et al. (2001) for a sample of
38 $S\ge 5$~mJy sources observed in two regions of the sky
corresponding to the ELAIS N2 and Lockman-Hole East fields. Although such
measurements are dominated by noise due to small-number statistics, it is
nevertheless interesting to note that -- as illustrated by Figure
\ref{fig:data} -- our predictions show full consistency with the data
(kindly provided by S. Scott).

\begin{figure}
\vspace{7cm}  
\includegraphics{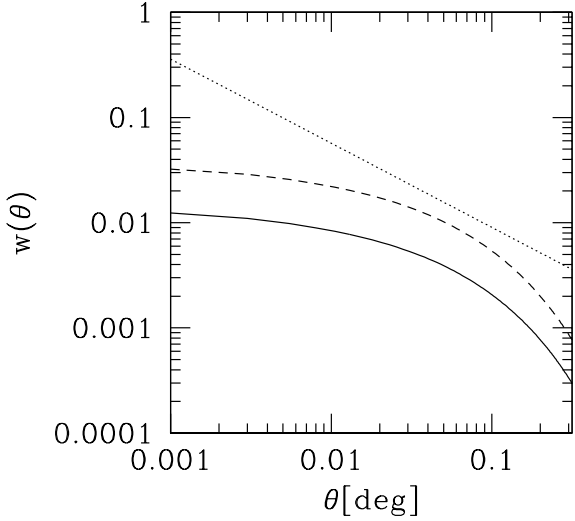}
\caption{Angular clustering of
unresolved ($S< 2$~mJy) sources at $850\,\mu$m. The solid curve is obtained for
$M_{\rm halo}/M_{\rm sph}=10$, while the dashed one represents the
case $M_{\rm halo}/M_{\rm sph}=100$. The dotted line corresponds
to $w(\theta)=\left(\theta/1''\right)^{-0.8}$ (Peacock et al.
2000). \label{fig:peacock}}
\end{figure}

Peacock et al. (2000) analysed the contribution of clustering of
unresolved ($S< 2$~mJy) SCUBA sources to the 850~$\mu$m background
fluctuations detected in the Hubble Deep Field (HDF) North. They
found some evidence for clustering of the background source
population, consistent with an angular correlation function of the
form $w(\theta)=\left(\theta/\theta_0\right)^{-\epsilon}$, with
$\epsilon = 0.8$ and $\theta_0=1''$. Figure \ref{fig:peacock}
presents our model predictions for the angular correlation
function $w(\theta)$ derived in the case of SCUBA galaxies with
$S< 2$~mJy; the redshift distribution spans the range $0.7\simlt
z\simlt 6$.


Our results are fully consistent with the constraints on structure
in the sub-millimetre background set by Peacock et al. (2000) and
not far from their preferred clustering model (dotted line in
Fig.~\ref{fig:peacock}), if $M_{\rm halo}/M_{\rm sph}=100$. The
shallower slope predicted by our model at small angles
($\theta\simlt 4\times 10^{-2}$ degrees) ultimately stems from the
fact that at high $z$ we enter the regime of linear growth of
density perturbations. This entails a flattening of the slope of
the two-point spatial correlation function $\xi(r,z)$ for $z
\simgt 2$. As a consequence, the angular correlation function
$w(\theta)$ obtained as in Eq.~(\ref{eq:limber}) by projecting
$\xi(r,z)$ over a wide range of redshifts, up to $z\simeq 6$, will
mirror this tendency to flatten out. Since a redshift range
$1\simlt z \simlt 6$ is also indicated by the analysis of Peacock
et al. (2000), the adoption of a flatter slope for $w(\theta)$ in
their clustering analysis might have been more appropriate.

Strong clustering substantially enhances cell-to-cell fluctuations
of the surface density of detected sources. The angular
correlation function $w(\theta)$ is in fact related to the second
moment (variance) of the galaxy distribution function via the
expression (Peebles 1980; Roche \& Eales 1999)

\begin{eqnarray}
\mu_2=\bar{N}+\left(\frac{\bar{N}}{\omega}\right)^2\Sigma^2,
\label{eq:mu2}
\end{eqnarray}
where $\bar{N}$ -- mean count of sources in the solid angle
$\omega$ -- represents the Poisson noise arising from the discrete
nature of the objects, and the normalised variance $\Sigma^2$ can
be written as

\begin{eqnarray}
\Sigma^2=\int w(\theta)\;d\omega_1\;d\omega_2.
\label{eq:sigma}
\end{eqnarray}
In the case of square cells with
$\omega=\Theta \times\Theta$, Eq.~(\ref{eq:sigma}) can be
expressed as the two-dimensional integral

\begin{eqnarray}
\Sigma^2(\Theta)=\Theta^2\int_0^\Theta dx\int_0^\Theta
w(\theta)\;dy, \label{eq:sigmaeasy}
\end{eqnarray}
with $\theta=\sqrt{x^2+y^2}$. \\

\begin{figure}
\vspace{7cm}  
\includegraphics{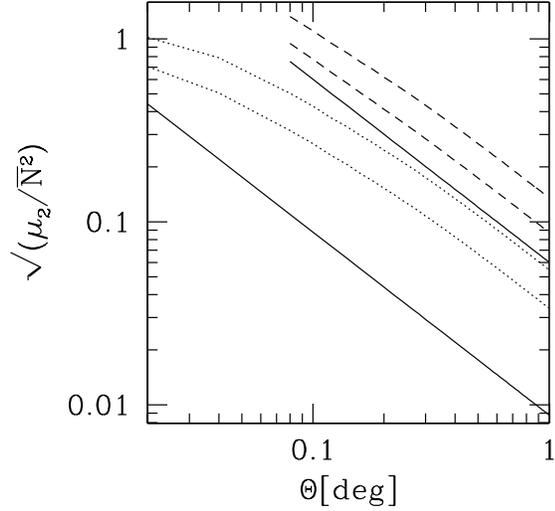} \caption{Predictions for the rms fluctuations relative
to the mean number of detected sources per cell of area
$\omega=\Theta^2$, $\mu_2^{1/2}/\bar{N}$, at 850~$\mu$m. Dashed
and dotted lines are for sources brighter than 10~mJy or 1~mJy,
respectively. Higher curves of each type correspond to $M_{\rm
halo}/M_{\rm sph}=100$, lower ones to $M_{\rm halo}/M_{\rm
sph}=10$. The solid lines show the Poisson
contributions for the two flux limits. The lines are drawn only
for angular scales such that the mean number of source in the
solid angle $\omega=\Theta^2$ is $> 1$. \label{fig:provacells}}
\end{figure}

The quantity in Eq.~(\ref{eq:sigmaeasy}) has been evaluated for
the three cases of low-, intermediate- and high-mass SCUBA
galaxies, introduced in this Section and based on the Granato et
al. (2001) model, and for the usual two different values of the
$M_{\rm halo}/M_{\rm sph}$ ratio. Results for the rms fluctuations
relative to the mean number of detected sources per cell,
$\mu_2^{1/2}/\bar{N}$ [see Eq.~(\ref{eq:mu2})] are presented in
Fig.~\ref{fig:provacells}, where dashed and dotted curves
correspond, respectively, to sources brighter than 10 and 1~mJy.
We have adopted a mean surface density per square degree of
$\bar{N}=180$ for a flux limit of 10~mJy, (see Scott et al. 2001),
and of $\bar{N}=1.3\times 10^4$ for a flux limit of 1~mJy (Granato
et al. 2001). Fig.~\ref{fig:provacells} clearly shows that the
contributions due to clustering are generally larger than the
Poisson contributions, shown by the two long-short dashed lines,
and in some cases by a large factor. For surveys covering small
areas of the sky the cell-to-cell fluctuations are large (up to
$\sim 100\%$). We therefore conclude that the high clustering
amplitude predicted for SCUBA sources can easily account for the
differences found in the number counts observed in different sky
regions by different groups (see e.g. Smail et al., 1997; Hughes
et al., 1998; Blain et al., 1999b; Barger et al., 1999a; Eales et
al., 2000; Scott et al., 2001).



\section{Multi-wavelength analysis of the contribution of clustering
to background fluctuations}


\begin{figure*}
\vspace{20.5cm}  
\includegraphics{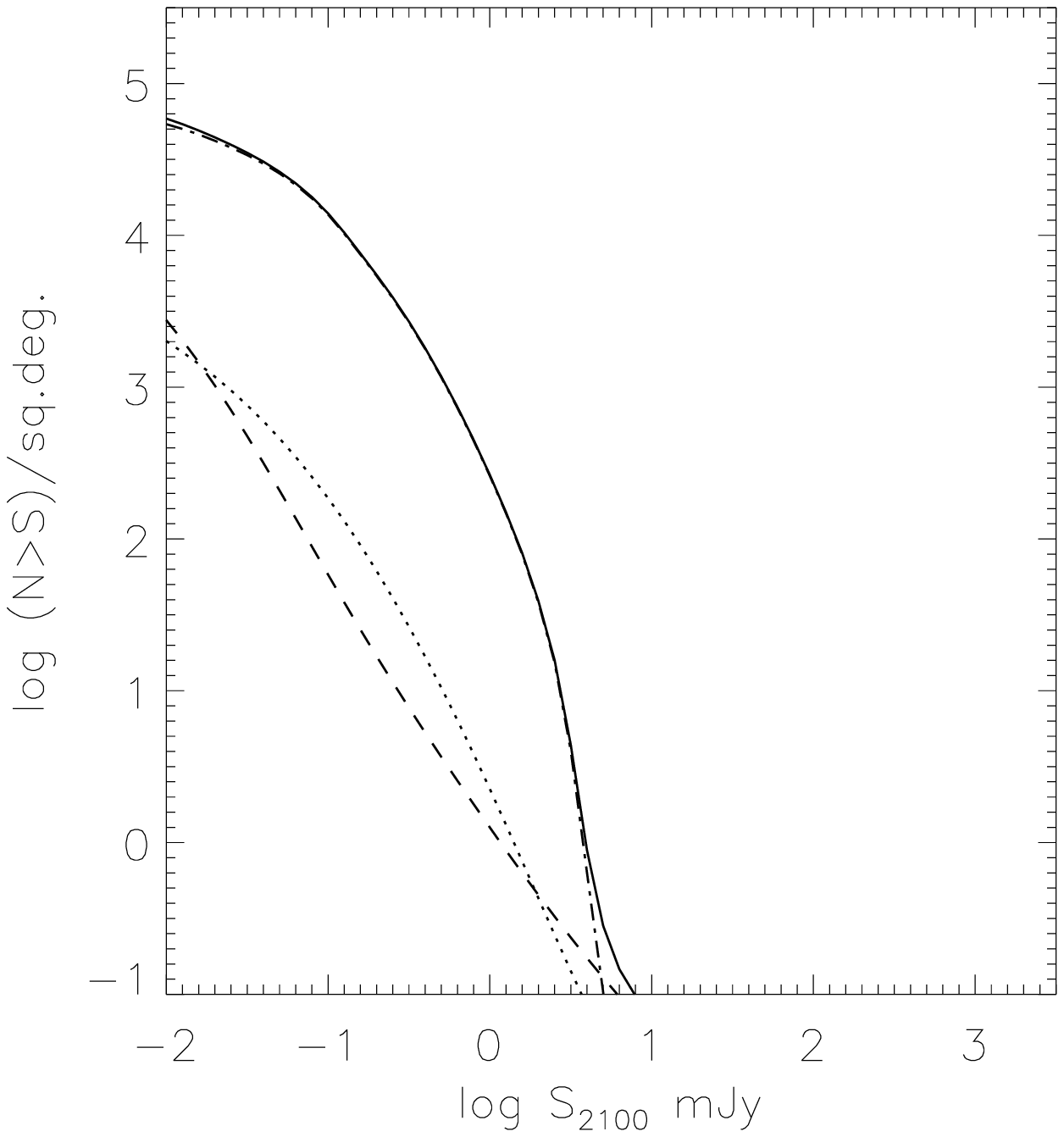} \includegraphics{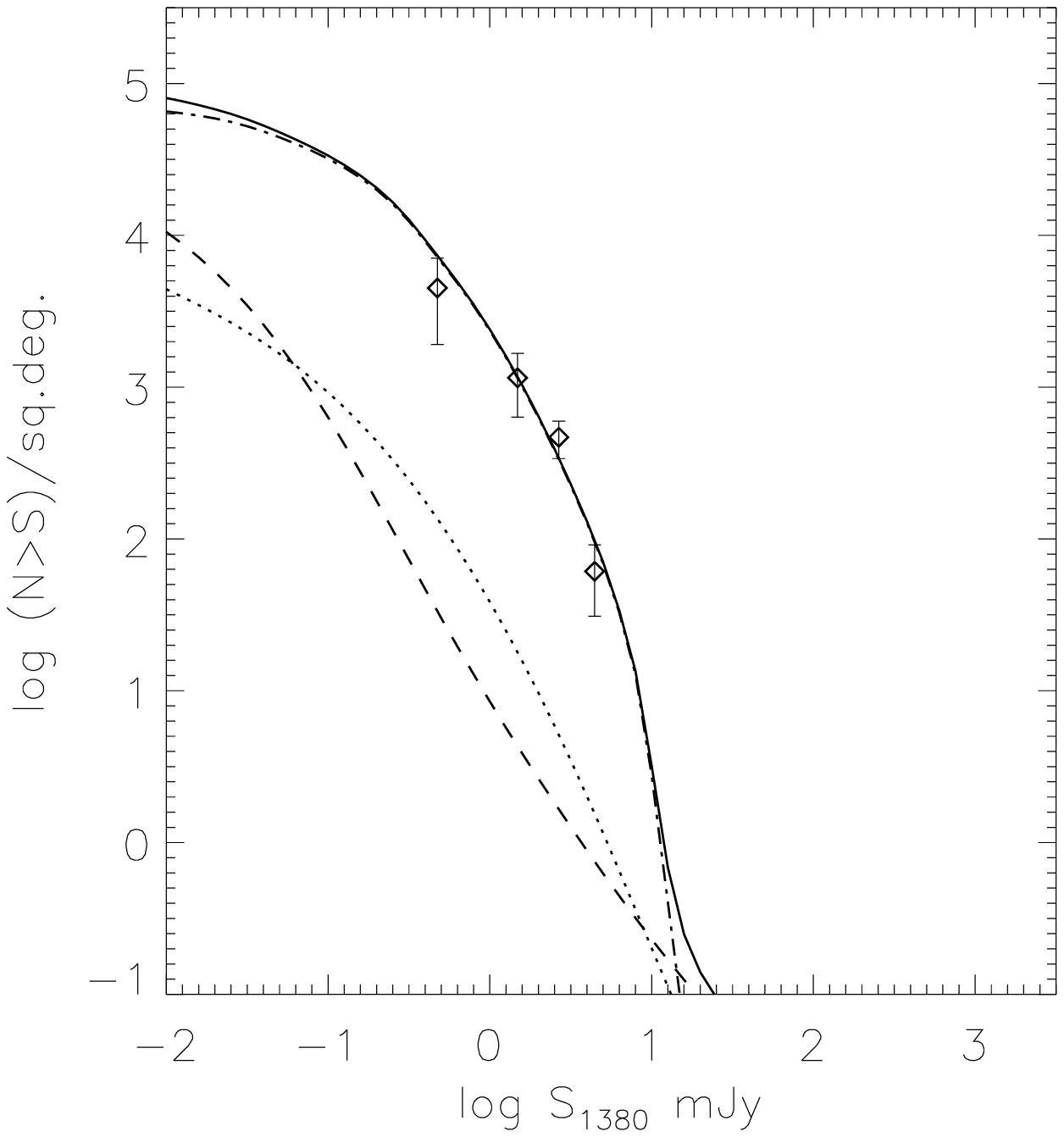} \includegraphics{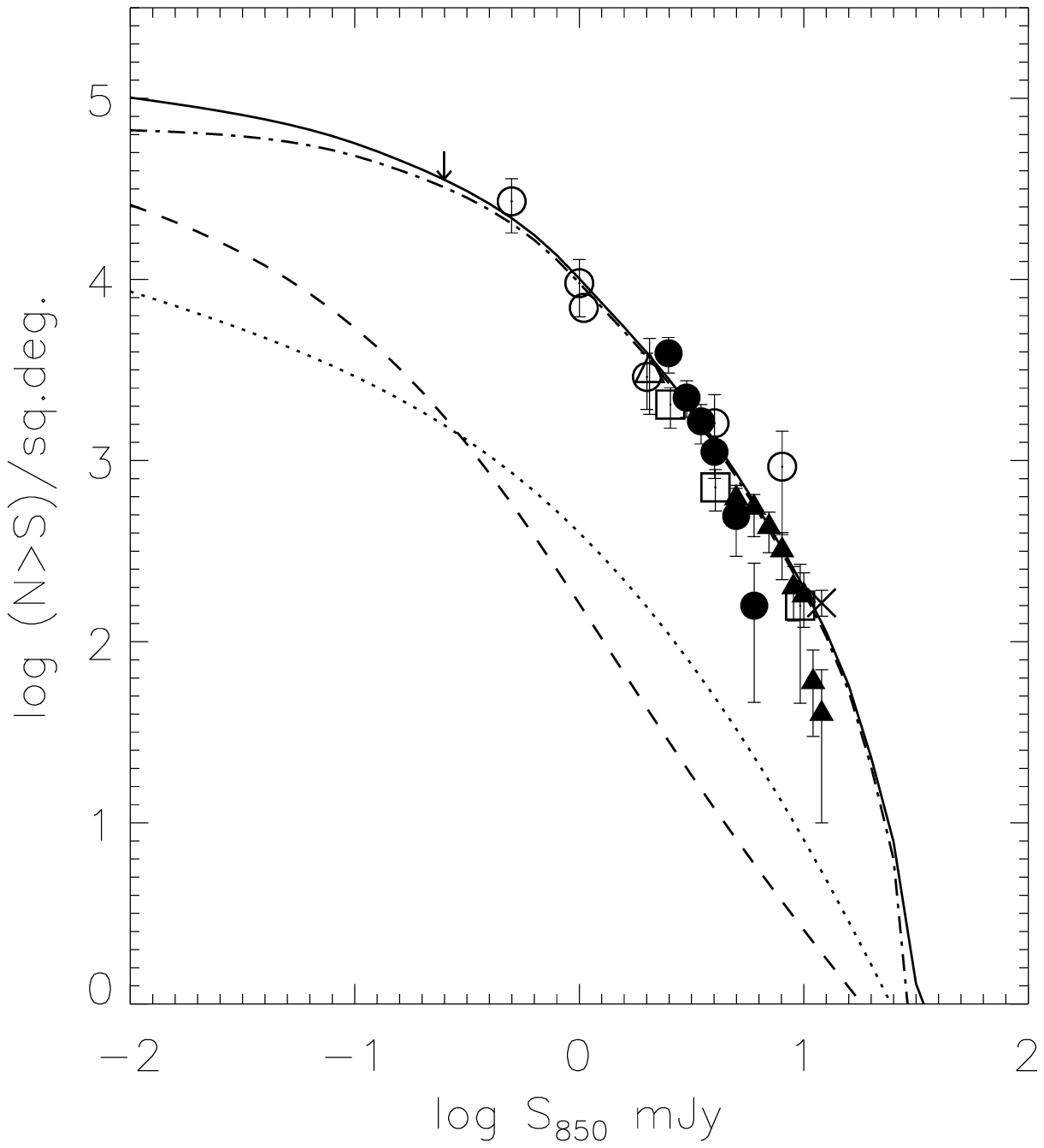} \includegraphics{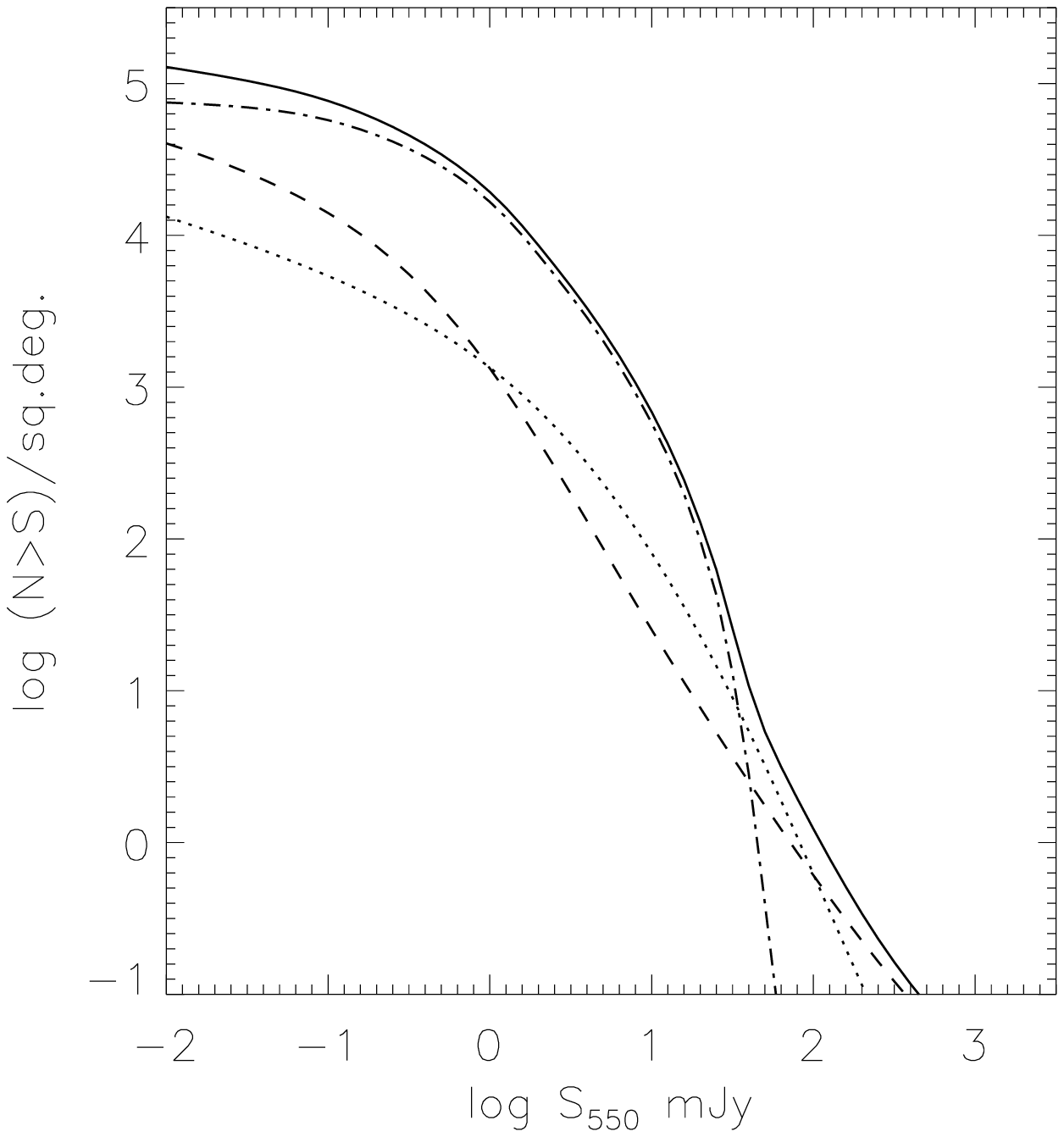}
\includegraphics{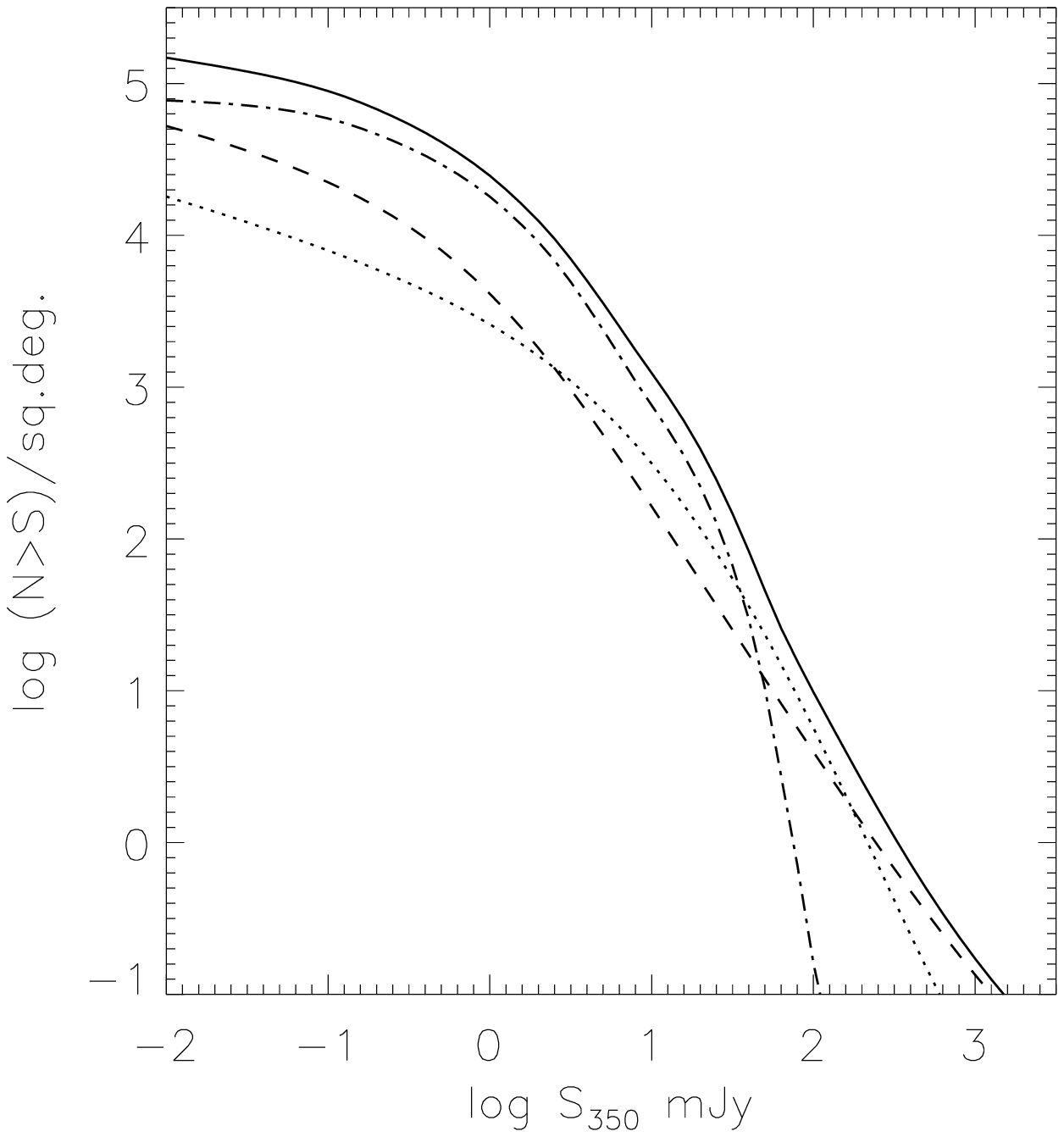} \includegraphics{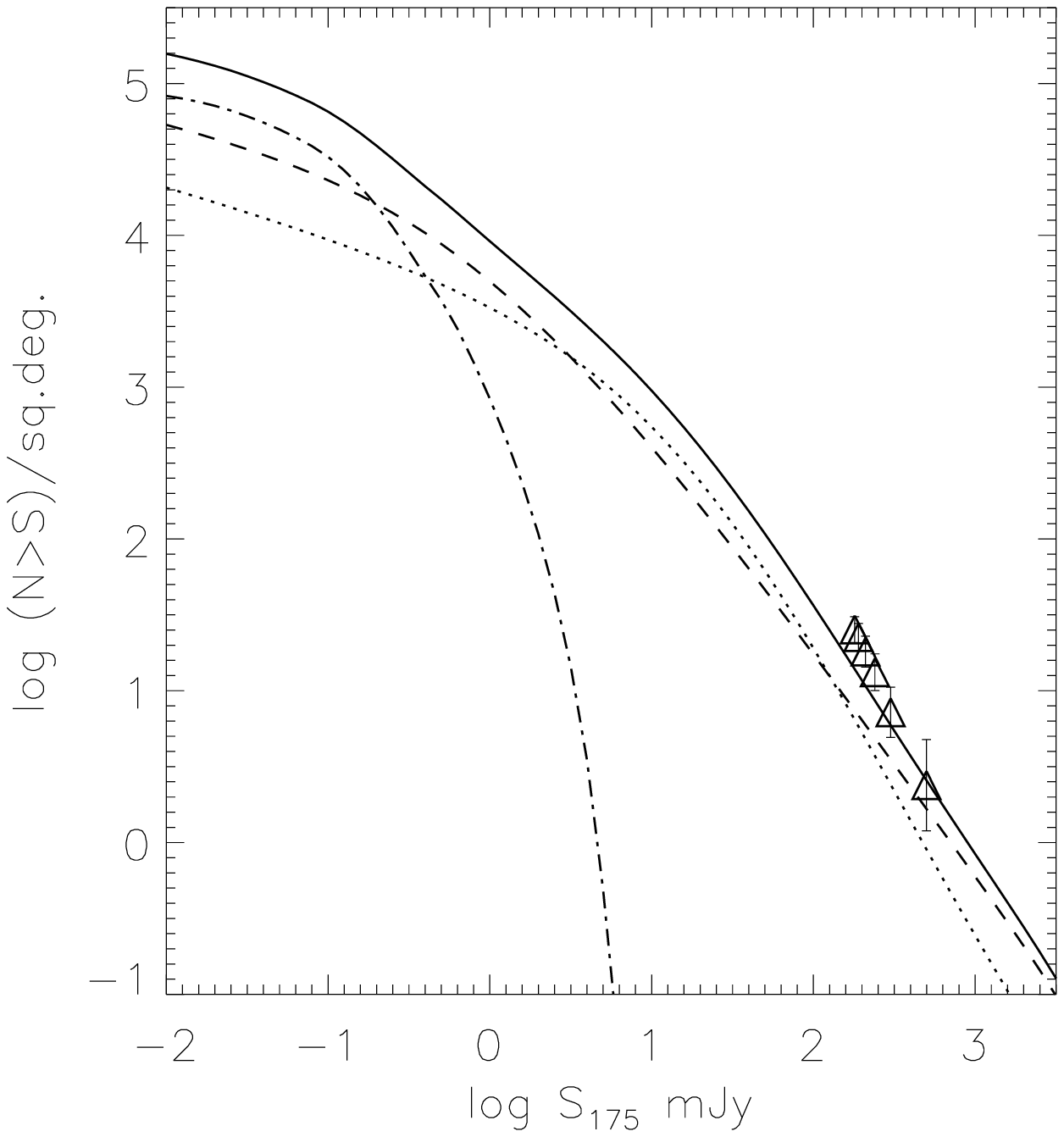} \caption{Integral number counts of galaxies
predicted by the model of Granato et al. (2001) at
$\lambda=2100\mu$m (top left-hand panel), $\lambda=1380\mu$m (top
right-hand panel), $\lambda=850\mu$m (middle left-hand panel),
$\lambda=550\mu$m (middle right-hand panel), $\lambda=350\mu$m
(bottom left-hand panel) and $\lambda=175\mu$m (bottom left-hand
panel). The solid lines show the total counts while the
dot-dashed, dashed, and dotted lines are respectively obtained for
the three populations of forming spheroidals, spirals and
star-burst galaxies. Data in the $1380\mu$m panel are derived from
MAMBO observations (Bertoldi et al., 2001); a flux density ratio
$S_{1200}/S_{1380}=1.5$ has been adopted to extrapolate to
$\lambda =1.38\,$mm the MAMBO flux densities at $\lambda =1.2\,$mm
(according to the model, such ratio varies between 1.4 and 1.6,
depending on the details of the dust temperature distribution).
Data at $175\mu$m are from Dole et al. (2001). Data at $850\mu$m
are from Blain et al. (1999b) (open circles), Hughes et al.
(1998)(open triangles), Barger et al. (1999) (open squares), Eales
et al. (2000) (filled circles), Scott et al. (2001) (filled
triangles), Borys et al. (2001) (crosses) \label{fig:counts}}
\end{figure*}

\begin{figure*}
\vspace{16cm}  
\includegraphics{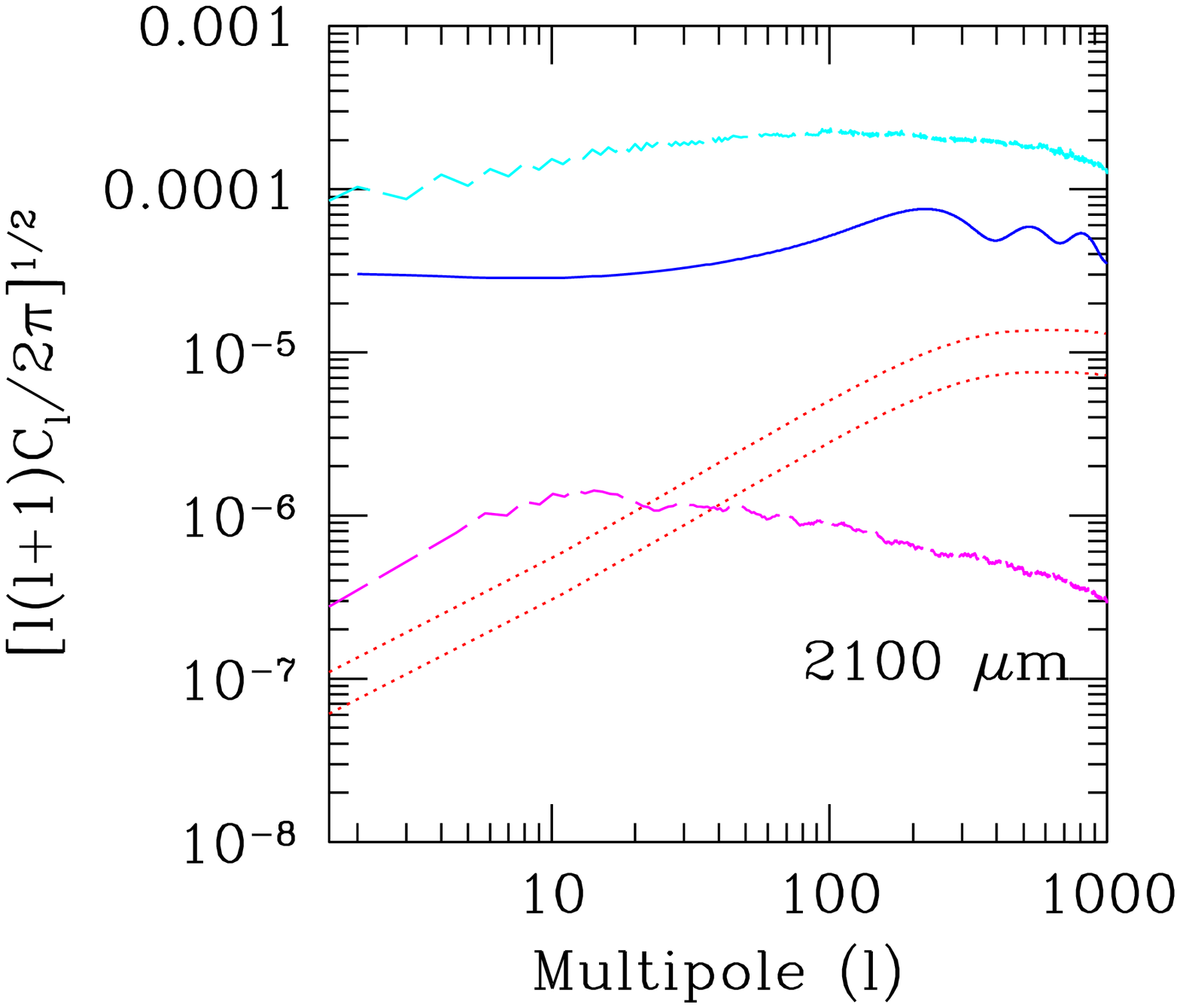} \includegraphics{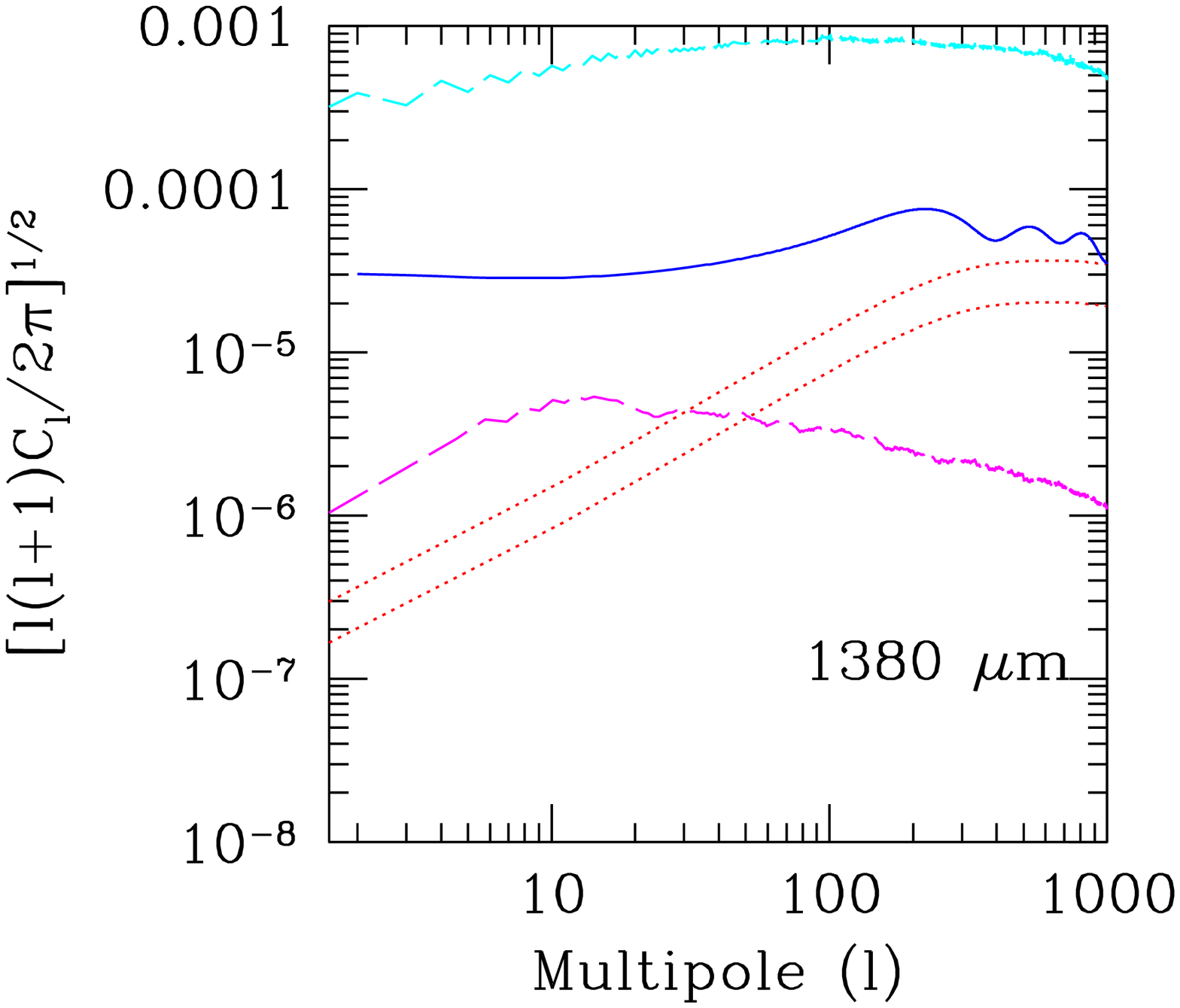}
\includegraphics{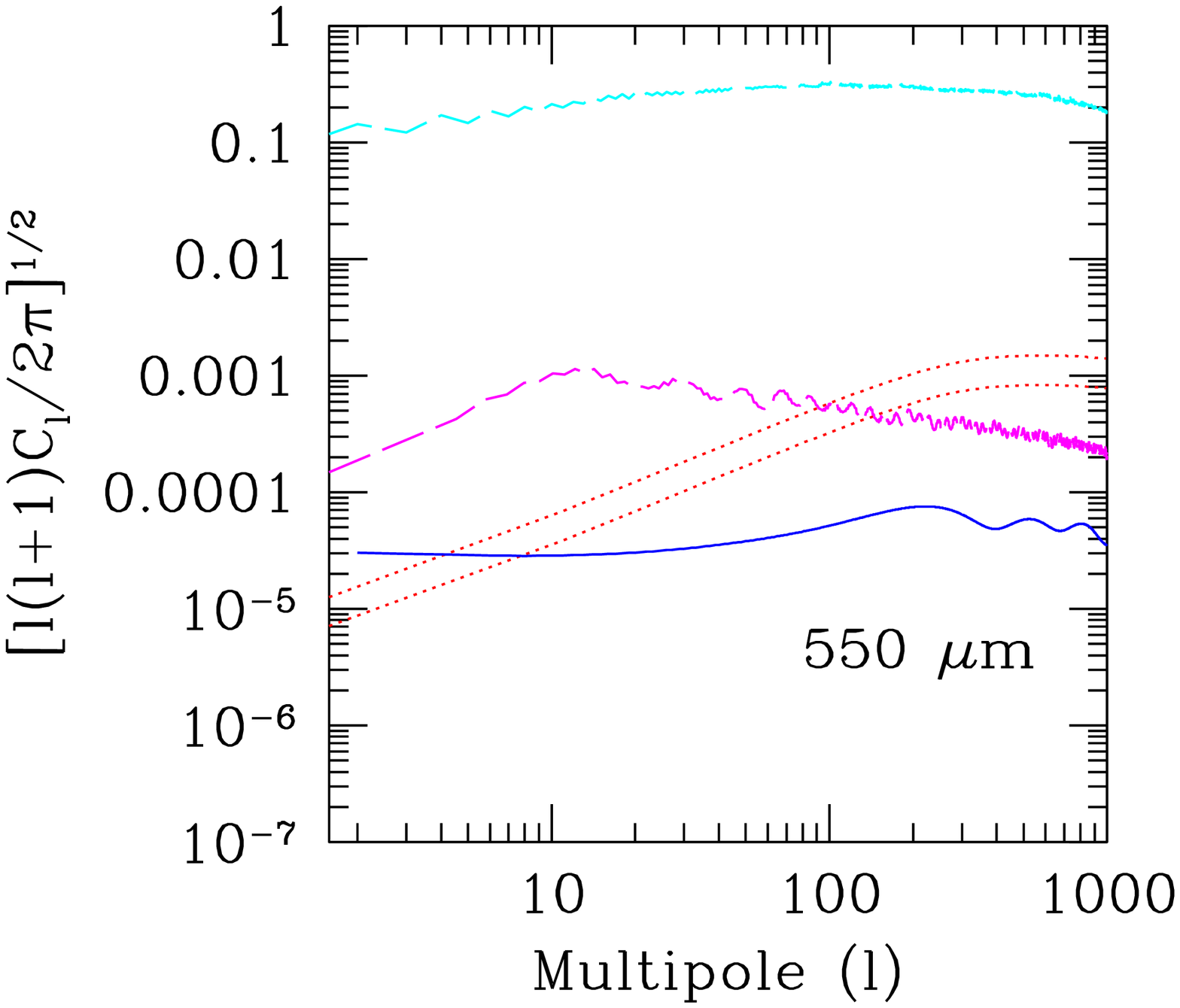}
\includegraphics{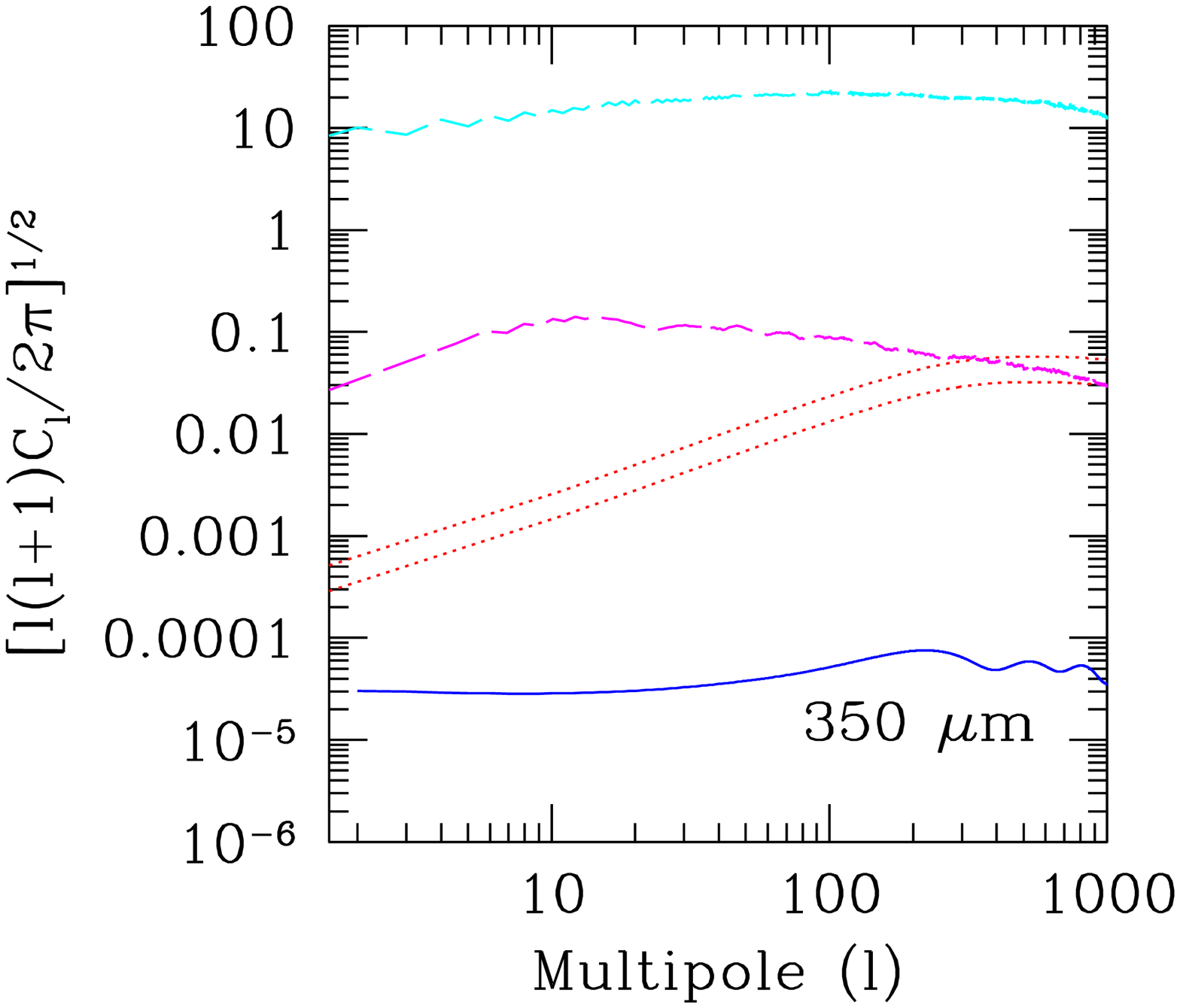} \caption{Predicted power spectrum
of temperature fluctuations $\delta T_l=\sqrt{l(l+1)C_l/2\pi}$ (in
units of K) as a function of the multipole $l$ at 2100~$\mu$m
(top-left panel), 1380~$\mu$m (top-right panel), 550~$\mu$m
(bottom-left panel) and 350~$\mu$m (bottom-right panel), central
frequencies of the  {\sc Planck}/HFI channels. The dotted curves
are for the 5$\sigma$ flux detection limit of each channel.
Higher curves of the same kind are for
$M_{\rm halo}/M_{\rm sph}$=100, lower ones for $M_{\rm
halo}/M_{\rm sph}$=10. The solid lines represent the power
spectrum of primary CMB anisotropies as predicted by a standard
$\Lambda$CDM cosmology ($\Lambda=0.7$, $\Omega_0=0.3$, $h_0=0.7$).
The upper long-dashed curves represent the contribution from
Galactic dust emission averaged all over the sky, while the lower
ones correspond to high galactic latitudes ($|b|\ge 80^\circ$)
only. \label{fig:fluct}}
\end{figure*}

The strong clustering of high-z forming spheroidal galaxies adds
an important contribution to background fluctuations. Measurements
of such a contribution can be informative on both the nature and the
properties of sources below the detection limit. On the other
hand, these fluctuations may have a significant impact on the
detectability of Cosmic Microwave Background (CMB) fluctuations.

MA2001 (but see also Scott \& White 1999 and Haiman \& Knox 2000)
estimated the power spectrum of clustering fluctuations at
850~$\mu$m. In the present Section we will apply the same analysis
to different wavelengths, corresponding to the effective
frequencies of the channels of the High Frequency Instrument (HFI)
of the {\sc Planck} ESA mission: $2100\,\mu$m $\rightarrow$ 143
GHz; $1380\,\mu$m $\rightarrow$ 217 GHz; $550\,\mu$m $\rightarrow$
545 GHz; $350\,\mu$m $\rightarrow$ 857 GHz. We will also work out predictions
at $\lambda=170 \mu$m, the wavelength probed by
the FIRBACK deep survey (Dole et al. 2001).

The counts of spheroidal galaxies are obtained from the model by
Granato et al. (2001). The counts of spiral and of star-burst
galaxies have been estimated by extrapolating to each considered
wavelength the $60\,\mu$m local luminosity functions of the two
populations, derived by Saunders et al. (1990), and evolving them
in luminosity as $L(z)= L(0) (1+z)^{\alpha}$, with $\alpha=3.5$ in
the case of starbursts and $\alpha=0$ for spirals. The
extrapolation in frequency has been carried out adopting the
observed spectral energy distributions of NGC6946 and NGC6090,
respectively in the case of spirals and star-burst galaxies.

The contribution to the counts of forming spheroidal galaxies,
when compared to the case for spirals and star-burst galaxies, is
found to decrease with decreasing wavelength because of the
increasingly less favourable K-corrections associated with the
former population. Spirals and starbursts consequently become more
and more important at shorter and shorter wavelengths, and
eventually dominate the counts for $\lambda\simlt 200 \mu$m. This
is illustrated by Fig.~\ref{fig:counts}, where we show the
contribution to the integral number counts (represented by the
solid curves) of the three populations of spheroidal galaxies
(dot-dashed lines), spirals (dashed lines) and starbursts (dotted
lines) at different wavelengths.

Figure~\ref{fig:counts} shows that -- under our assumptions --
star-bursts and spirals constitute a negligible fraction of the
total number of sources at mm/sub-mm wavelengths, the dominant
contribution coming from high-z forming spheroidal galaxies (the
same which -- according to Granato et al. (2001) -- show up as
SCUBA sources at $\lambda=850\mu$m), at all but very bright fluxes
(with a flux threshold which varies with wavelength) where, as
already discussed in the introduction and extensively discussed in
Section 4, the number counts of spheroidal galaxies experience an
exponential decline. The situation is reversed at 175~$\mu$m
(bottom right panel of Figure~\ref{fig:counts}), where at all but
very faint fluxes, spirals and star-bursts dominate the predicted
number counts. In the following analysis we will therefore only
consider the population of spheroidal galaxies when dealing with
the mm/sub-mm wavelength range.

\begin{figure}
\vspace{8cm} \includegraphics{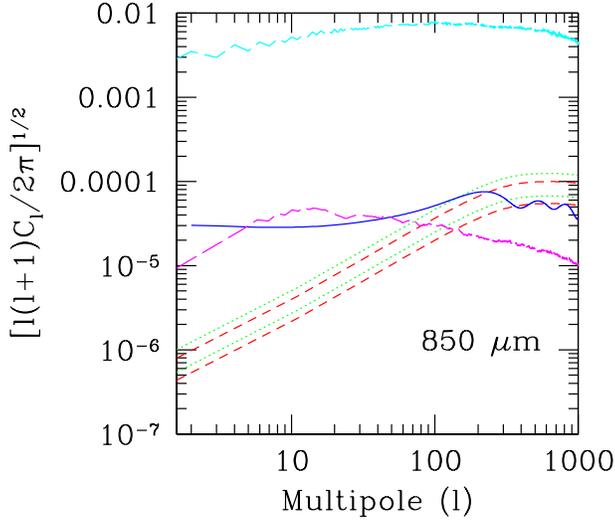} \caption{As in Figure
\ref{fig:fluct}, but for the 850 $\mu$m case. Dotted lines are for
a detection limit $S_d=100$~mJy while dashed ones for
$S_d=10$~mJy. Higher curves of the same kind are for $M_{\rm
halo}/M_{\rm sph}$=100, lower ones for $M_{\rm halo}/M_{\rm
sph}$=10.\label{fig:dT}}
\end{figure}

The angular correlation function of intensity fluctuations due to
inhomogeneities in the space distribution of unresolved sources
(i.e. with fluxes fainter than some threshold $S_d$), $C(\theta)$,
can be expressed as the sum of two terms $C_P$ and $C_C$, the
first one due to Poisson noise (fluctuations given by randomly
distributed objects), and the second one owing to source
clustering (see MA2001 and De Zotti et al. 1996 for a detailed
discussion). In the case of highly clustered sources such as
forming spheroidal galaxies, the Poisson term $C_P$ is found (see
e.g. Scott \& White 1999) to be negligible when compared to $C_C$.
We can therefore safely assume $C\simeq C_C$, whose expression is
given by
\begin{eqnarray}
C_C(\theta)&=&\left({1\over 4\pi}\right)^2  \int_{z_{(L_{\rm
min},S_d})}^{z_{\rm max}}\!\!\!\!\!\!\!\!\! dz\;b_{\rm
eff}^2(M_{\rm min},z)\;
\frac{j^2_{\rm eff}(z)} {(1+z)^4}\left(\frac{dx}{dz}\right)^2\nonumber\\
&\cdot& \int_0^\infty d(\delta z)\;\xi(r,z), \label{eq:cth}
\end{eqnarray}
with $b_{\rm eff}(M_{\rm min},z)$ defined by
Eq.~(\ref{eq:bias}), $x$ being the comoving radial coordinate,
$\delta z=c/H_0\;u$
($u$ is introduced in Eq. (1) and $c$ is the speed of light), $z_{\rm max}$
the redshift when the sources begin to shine and $z(S_d,L)$ the
redshift at which a source of luminosity $L$ is seen with a flux
equal to the detection limit $S_d$. The effective volume
emissivity $j_{\rm eff}$ is expressed as:
\begin{eqnarray}
j_{\rm eff}=\int_{L_{\rm min}}^{{\rm min}[L_{{\rm
max},L(S_d,z)}]}\!\!\!\!\!  \Phi(L,z)\; K(L,z)\; L\;d{\rm log}L,
\label{eq:jeff}
\end{eqnarray}
$\Phi(L,z)$ being the luminosity function (per unit logarithmic
luminosity interval), $K(L,z)$ the K-correction and $L_{\rm max}$
and $L_{\rm min}$ respectively maximum and minimum local
luminosity of the sources.

$C_C(\theta)$ in Eq.~(\ref{eq:cth}) has been evaluated separately
for the three cases of low-, intermediate- and high-mass objects,
by plugging in Eq.~(\ref{eq:jeff}) the appropriate luminosity
functions (Granato et al. 2001). The total contribution of
clustering to intensity fluctuations has then been derived by
adding up all the values of $C_C(\theta)$ obtained for the
different mass intervals and by also taking into account the
cross-correlation terms between objects of different masses,
according to the expression (where we have dropped the index $C$)
\begin{eqnarray}
C^{TOT}(\theta)=\sum_{i,j=1}^3 {\sqrt {C_i(\theta)C_j(\theta)}}.
\label{eq:ctot}
\end{eqnarray}
The angular power spectrum of the intensity fluctuations can then
be obtained as
\begin{eqnarray}
C_l=\langle |a_l^0|^2 \rangle=\int_0^{2\pi}\int_0^\pi[\delta
T(\theta)]^2\;P_l(\rm{cos}\theta)\;\rm{sin}\theta\;d\theta\;d\phi,
\end{eqnarray}
with
\begin{eqnarray}
\delta T(\theta)&=&\langle\left(\Delta T\right)^2 \rangle^{1/2} =
\frac{\lambda^2\! \sqrt{C^{TOT}(\theta)}}{2\;k_b}\left[\exp\left
(\frac{h\nu}{k_b T}\right)-1 \right]^2 \nonumber\\
&\cdot& \exp\left(- \frac{h\nu}{k_b
T}\right)/\left(\frac{h\nu}{k_b T} \right)^2,
\end{eqnarray}
which relates intensity and brightness temperature fluctuations.

This analysis has been performed for the different central
frequencies of the {\sc Planck} HFI, and Figure \ref{fig:fluct}
shows the results in terms of $\delta T_l=\sqrt{l(l+1)C_l/2\pi}$
(in units of K) at 2100~$\mu$m, 1380~$\mu$m, 550~$\mu$m and
350~$\mu$m. In each of the four panels of Fig.~\ref{fig:fluct},
dotted curves are obtained by adopting the 5$\sigma$ detection
limits of the different {\sc Planck} channels ($S_{\rm
lim}=300$~mJy at 2100~$\mu$m; $S_{\rm lim}=200$~mJy at
1380~$\mu$m; $S_{\rm lim}=450$~mJy at 550~$\mu$m; $S_{\rm
lim}=700$~mJy at 350~$\mu$m) estimated by Toffolatti et al.
(1998). In Figure \ref{fig:dT} we also
plot predictions for temperature fluctuations at  850 $\mu$m, where the
contribution from star-forming spheroidals has been corrected for a mistake
in MA2001. As in the former cases, dotted curves are obtained for a
5$\sigma$ detection limit of the relevant {\sc Planck} channel, corresponding
to $S_{\rm lim}=100$~mJy, while dashed ones illustrate the case for ten 
times higher sensitivity ($S_d=0.1\;S_{\rm lim}$). 

The solid curves in Figs.~\ref{fig:fluct} and \ref{fig:dT}
show, for comparison, the
power spectrum of primary Cosmic Microwave Background (CMB)
anisotropies corresponding to a flat $\Lambda$CDM cosmology,
calculated with the CMBFAST code developed by Seljak \&
Zaldarriaga (1996). The relative importance of fluctuations
due to clustering rapidly increases with decreasing wavelength.
CMB anisotropies on small angular scales are exceeded at wavelengths
$\lambda\le 850\mu$m. This high clustering signal mostly comes from
massive galaxies with bright fluxes, which lie at substantial
redshifts and are therefore highly biased tracers of the
underlying mass distribution.  Also the negative K-corrections
increase their contribution to the effective volume emissivity
[Eq.~(\ref{eq:jeff})] and therefore to the fluctuations.

This implies that important information on the clustering
properties of faint sub-mm/far-IR galaxies (and hence on physical
properties such as their mass and/or the amount of baryons
involved in the star-formation process) will reside in the {\sc
Planck} maps at frequencies greater than 353 GHz where, however,
the dominant signal is expected to come from interstellar dust
emission.

In order to quantify this last effect, we have calculated the
expected contribution of Galactic dust emission to background
fluctuations. To this end, the dust emission maps constructed by
Schlegel et al. (1998) by combining together IRAS and DIRBE data,
have been extrapolated to all relevant frequencies assuming a
grey-body spectrum with emissivity index $m=2$ and average dust
temperature $T_{\rm DUST}=18$~K.

Fig.~\ref{fig:fluct} shows the power spectrum of Galactic dust
emission averaged all over the sky (upper long-dashed curves).
This signal dominates at all wavelengths due to contribution from
the Galactic plane, highly contaminated by dust. Nevertheless, it
is still possible to detect the clustering signal of sub-mm
galaxies (and also recover the CMB power spectrum for $\lambda
\simgt 10^3 \mu$m) if one restricts the analysis to high enough
Galactic latitude regions (i.e. $|b|\ge 80^\circ$, lower
long-dashed curves), which are least affected by dust emission.

It turns out that, both for $\lambda=850\mu$m (see also
Magliocchetti et al. 2001b) and $\lambda=550\mu$m, the dust
contribution in these regions becomes less important than the one
due to the clustering of unresolved sources at any $l\simgt 100$.
For $\lambda=350\mu$m the signal due to Galactic dust is instead
found to hide all the other sources of background fluctuations at
all Galactic latitudes, except possibly in the case of very high
($l\simgt 800$) multipoles.

Background fluctuations due to unresolved extragalactic sources
have recently been measured by Lagache \& Puget (2000) at
$170\,\mu$m in a field covered by the FIRBACK survey with ISOPHOT
(Dole et al. 2001).

As already discussed in this Section, the composition of counts at
such relatively short wavelengths is substantially different from
those considered so far because of the dominant presence of
low-to-intermediate-redshift spiral and starburst galaxies.
Calculations of the background fluctuations due to clustering of
unresolved sources have then to take into account these two
classes of sources, together with the population of spheroidal
galaxies.

We use again Eqs.~(\ref{eq:cth}) and (\ref{eq:jeff}) but, when
dealing with
 spiral and starburst galaxies, an important difference in the form of the
the bias factor $b_{\rm eff}(M_{\rm min},z)$ appearing in
Eq.~(\ref{eq:cth}) has to be taken into account. In this case, in
fact, $b_{\rm eff}(M_{\rm min},z)$ assumes the form:
\begin{eqnarray}
b(z)=1+\frac{b_0-1}{D(z)} \label{eq:biassp}
\end{eqnarray}
(Fry 1996), independent of the mass of the haloes hosting such
sources and merely a function of redshift -- via the linear growth
rate $D(z)$ -- and of the local bias $b_0\equiv
b(z=0)=\sigma_{8,gal}/\sigma_8$, where $\sigma_{8,gal}$ is the rms
fluctuation amplitude in a sphere of radius 8 h$^{-1}$Mpc as
measured in the local universe for each population. As a
consequence, no distinction between low-, intermediate- and
high-mass objects is required.

The choice for a bias evolution of the form given by
Eq.~(\ref{eq:biassp}) stems from a number of results (see e.g.
Baugh et al. 1999; Magliocchetti et al. 1999; Magliocchetti et al.
2000, to mention just a few) showing that bias models such as the
one adopted for high-z forming spheroidal galaxies and illustrated
by Eq.~(\ref{eq:bias}) do not provide a good fit for $z\simlt 1$,
where most spirals and star-burst galaxies with $170\,\mu$m fluxes lying
within the range relevant to this work, are found.

We have estimated the amplitudes $\sigma_{8,g}$ from the
clustering properties by spectral type determined by Loveday,
Tresse \& Maddox (1999) from Stromlo-APM survey data. We find
$\sigma_{8,g}=0.93$ for galaxies with weak emission lines
(spirals) and $\sigma_{8,g}=0.66$ for galaxies with strong
emission lines (starbursts). This leads to $b_0=0.93$ for spirals
and to $b_0=0.66$ for starbursts, implying this latter
population to be strongly anti-biased with respect to the underlying
mass distribution.

\begin{figure}
\vspace{7cm}  
\includegraphics{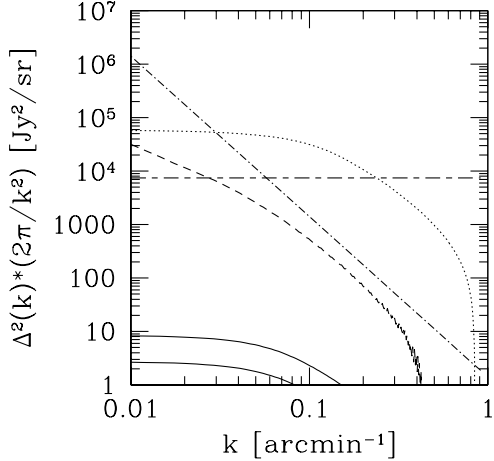}
\caption{Predictions for the power-spectrum of intensity
fluctuations due to clustering of galaxies fainter than a
detection limit $S_d=135$~mJy at $\lambda=170\mu$m. The solid
curves are for the population of spheroidal galaxies (the higher
one referring to $M_{\rm halo}/M_{\rm sph}=100$, the lower one to
$M_{\rm halo}/M_{\rm sph}=10$), while the dashed and dotted curves
respectively show the predictions for spiral and star-burst
galaxies. The short/long-dashed curve represents a white noise power
spectrum $P_P=7400$ Jy$^2$/sr, while the  dashed-dotted one illustrates the
cirrus confusion noise (both from Lagache \& Puget, 2000).
\label{fig:fluct175}}
\end{figure}

The contribution of clustering to $C(\theta)$ has then been calculated
for the three populations of spheroidals, spirals and starbursts.
The angular power-spectrum $\Delta^2(k)$, in the units used by
Lagache \& Puget (2000) (Jy$^2$~sr$^{-1}$), has been derived according to the
expression (see Peacock 1997):
\begin{eqnarray}
\Delta^2(k)=k^2\;\int_0^\infty
C(\theta)\;J_0(k\theta)\;\theta\;d\theta,
\label{eq:Pk}
\end{eqnarray}
where $k$ is the angular wavenumber and $J_0$ is the zero-th order
Bessel function. The results for a detection limit $S_d=135$~mJy,
corresponding to three times the confusion noise of the FIRBACK
survey (Dole et al. 2001), are shown in Fig.~\ref{fig:fluct175},
where the contribution of spheroidal galaxies is shown by the
solid curves, while the dashed and dotted curves respectively show
the predictions for spiral and starburst galaxies.

As expected, the contribution of spheroidal galaxies at this wavelength is
negligible when compared to those originating from the clustering of
both star-burst galaxies and spirals. Also, despite the strongly negative
bias,
the signal obtained for starbursts turns out to be stronger than the one
predicted for spiral galaxies. This is due to the different evolutionary
properties of the two populations, since no evolution is assumed for spiral
galaxies, while strong evolution is instead allowed for in the case of
starbursts (cf. beginning of this Section).

The short/long-dashed curve in Figure \ref{fig:fluct175}
represents the Poisson contribution $P_P=7400$ Jy$^2$/sr derived
from the Guiderdoni et al. (1997) predictions, while the
dashed-dotted curve illustrates the contribution to the observed
power-spectrum of the Galactic cirrus (Lagache \& Puget 2000).
According to our predictions, one then has that fluctuations
stemming from the clustering of starburst galaxies are detectable
above other sources of signal, and this happens in the range of
wavenumbers $0.04\simlt k\;({\rm arcmin}^{-1})\simlt 0.3$.

These findings, together with the predicted magnitude of the total
(non-Poissonian) power spectrum for extragalactic sources (practically
coincident with the dotted line in Figure \ref{fig:fluct175}) at 170 $\mu$m,
are fully consistent with the measurements of Lagache \& Puget (2000).
We therefore conclude that an excess of signal (with respect to Poisson
fluctuations) due to clustering of
unresolved/strongly evolving starburst galaxies, could possibly be
discerned in the 170 $\mu$m FIRBACK maps.

\section{Gravitational Lensing}
\label{strong}

\begin{figure}
\vspace{9.5cm} \includegraphics{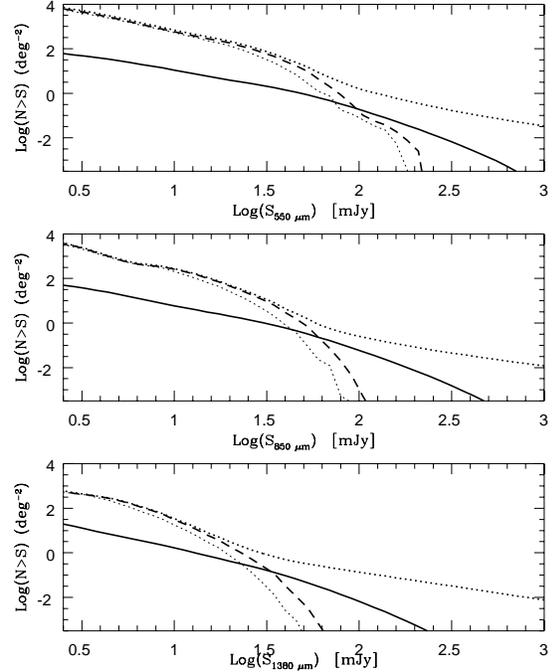} \caption{Integral source
counts per square degree for the model by Granato et al. (2001) at
550 (top), 850 (middle) and $1380\,\mu$m. The light dotted lines,
the heavy dashed lines, and the solid lines respectively refer to
the unlensed, weakly lensed, and strongly lensed counts of forming
spheroids only. The heavy dotted line shows the total counts (sum
of weakly and strongly lensed sources), including contributions
from populations whose counts are essentially unaffected by
lensing (such as ``flat''-spectrum radio sources), as described in the 
text.  }
\label{cg}
\end{figure}

\begin{figure}
\vspace{9.5cm} \includegraphics{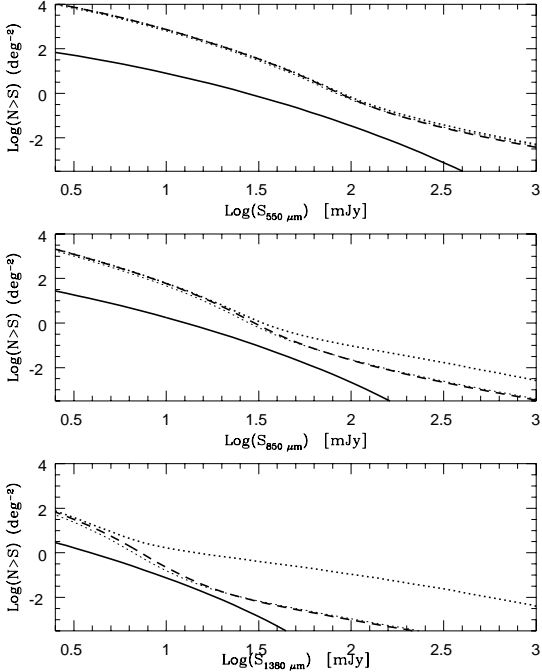} \caption{Integral source
counts for the model by Rowan-Robinson (2001). The meaning of the
lines is analogous to that in Fig.~8. In this case, however, the
light dotted, the heavy dashed (superimposed to the dotted), and solid 
lines include the contributions of all the populations considered by 
Rowan-Robinson.
The heavy dotted line also includes ``flat''-spectrum radio
sources (see text).
} \label{crr}
\end{figure}

\begin{figure}
\vspace{9.cm} \includegraphics{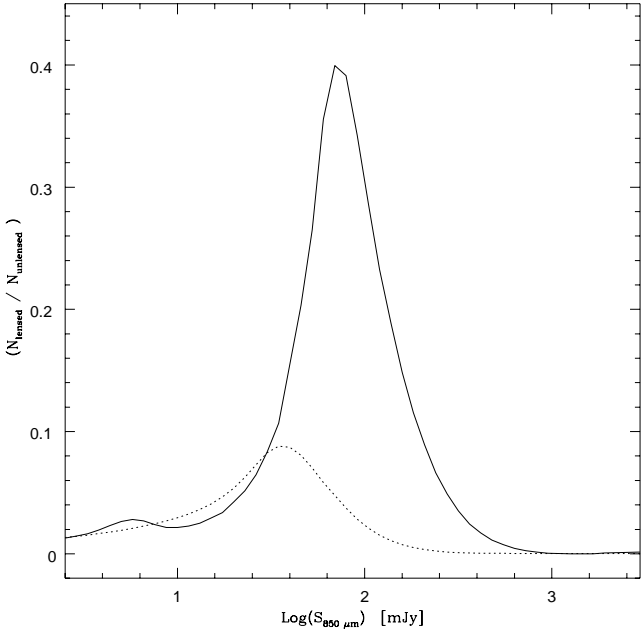} \caption{Ratio of lensed to unlensed
counts at 850 $\mu$m: the solid line refers to the model by
Granato et al. (2001), the dotted line to the one by
Rowan-Robinson (2001). 
}
\label{ratio}
\end{figure}

It is now well established (Franceschini et al. 1991, 1994,
2001; Blain \& Longair 1993;  Pearson \& Rowan-Robinson 1996;
Guiderdoni et al. 1997, 1998; Dwek et al. 1998; Blain et al.
1999b; Devriendt \& Guiderdoni 2000; Pearson et al. 2001; Takeuchi
et al. 2001; Rowan-Robinson 2001) that the coupling of the
strongly negative K-correction at mm/sub-millimetre wavelengths --
due to the steep increase with frequency of dust emission in
galaxies -- with the strong cosmological evolution demonstrated by
both ISO and SCUBA data (Elbaz et al. 1999; Smail et al. 1997;
Hughes et al. 1998; Barger et al. 1999a; Eales et al. 2000) and by
the intensity of the far-IR background (Puget et al. 1996; Fixsen
et al. 1998; Schlegel et al. 1998; Hauser et al. 1998; Lagache et
al. 1999, 2000), greatly emphasises high-redshift sources. This
yields very steep counts which maximise the magnification bias
(Peacock 1982; Turner et al. 1984) and lead to a high probability
for such sources to be gravitationally lensed (Peacock 1982;
Turner et al. 1984). As already discussed by Blain (1996, 1997,
1998a,b, 1999, 2000), this corresponds to a fraction of lensed
sources expected to show up in the mm/sub-millimetre band which is
much larger than what is found in surveys at other wavelengths. For
instance, Blain (1998a) predicts about 0.6 to 5\% of the point
sources observed in the future by the High Frequency Instrument
(HFI) on board of the ESA satellite P{\sc lanck} to be lensed.

A distinctive feature of the astrophysically grounded model by
Granato et al. (2001) is that both the lensing probability and the
magnification bias at mm/sub-mm wavelengths are substantially
higher than what other current, mostly
phenomenological, models imply which also successfully account for
SCUBA/MAMBO counts (Rowan-Robinson 2001; Takeuchi et al. 2001;
Pearson et al. 2001; Blain et al. 1999b). This is because,
according to this model, most galaxies detected in blank-field
SCUBA and MAMBO surveys are interpreted as massive ellipticals at
$z \simgt 2$, in the process of building their stellar populations
with very high star-formation rates (typically from a few hundreds
to $\sim 1000\,\hbox{M}_\odot\,\hbox{yr}^{-1}$). This leads to an
extremely steep, essentially exponential, decline of the bright
tail of mm/sub-mm counts of this galaxy population due to the fact
that, in this model, dust emission from these objects rapidly
fades away at $z\simlt 2$ when the bulk of their star formation is
essentially over (Cohen 2001). The bright counts therefore somehow
reflect the high luminosity tail of the luminosity function which,
in turn, reflects to some extent the fact that, in the
hierarchical clustering scenario, massive halos are exponentially
rare at high redshifts. Such a steep decrease of the number counts
for fluxes $10{\rm mJy} \simlt S_{\rm 850\mu m}\simlt 100{\rm
mJy}$ implies a large fraction of strongly lensed galaxies to
appear at bright mm/sub-mm fluxes.

We recall here the main aspects of our model for gravitational
lensing, referring to our previous work (Perrotta et al. 2001) for
details. In the next Section we present quantitative predictions
for the fraction of lensed spheroids to be compared with
forthcoming observations.

Lensing statistics is expressed by the probability for a source at redshift
$z_{\rm s}$ to be lensed with magnification $>\mu$: it is obtained by dividing
the total lensing cross section by the area of the source sphere as
\begin{eqnarray}
  P(\mu,z_{\rm s})&=&\frac{(1+z_{\rm s})^2}{4\pi r^2(z_{\rm s})}\,
  \int_0^{z_{\rm s}}\,{\rm d}z\,\frac{dV}{dz}\,(1+z)^3\nonumber\\
  &\times& \int{\rm d}M\,\sigma(\mu,z,z_{\rm s},M){dn\over dM}(z,M)\; ,
\label{Prob}
\end{eqnarray}
where $r(z)$ is the comoving radial distance to redshift $z$,
$dV/dz$ is the proper volume element per unit redshift, and
$dn(z,M)/dM$ the comoving number density of the lenses. Since we
are dealing with gravitational lensing by dark matter haloes, we
assume that the lens distribution follows the Sheth \& Tormen
(1999) mass function
\begin{eqnarray}
  \frac{{\rm d}n}{{\rm d}M} &=& \sqrt{\frac{2aA^2}{\pi}}\,
  \frac{\rho_0}{M^2}\,\frac{\delta_{\rm c}(z)}{\sigma(M)}\,
  \left[1+\left(\frac{\sigma(M)}{\sqrt{a}\delta_{\rm c}(z)}
  \right)^{2p}\right]\nonumber\\
  &\times& \left|\frac{{\rm d}\ln\sigma}{{\rm d}\ln M}\right|\,
  \exp\left(-\frac{a\delta_{\rm c}^2(z)}{2\sigma(M)^2}\right)\; ,
\label{Sheth}
\end{eqnarray}
where the best-fit values of the parameters for our cosmological
model are $a=0.707$, $p=0.3$, and $A\simeq0.3222$. $\rho_0$ is the
mean mass density at a reference epoch $t_0$, which we assume to
be the present time, and $\sigma^2$ is the variance of linear
density fluctuations at the present epoch, smoothed with a
spherical top-hat filter $W_R(k)$ enclosing a mass $M$.
$\delta_{\rm c}^2(z)$ is the linear density contrast of an object
virialising at $z$, linearly evolved to the present epoch.

Quite independent of the lens model, $P(\mu ,z)$ decreases as
$\mu^{-2}$ for $\mu\gg1$, hence the high magnification tail of the
probability per unit magnification can be written as $p(\mu
,z)=-dP(\mu ,z)/d\mu\propto a(z)\,\mu^{-3}$. Equation
(\ref{Sheth}) assumes non-overlapping cross sections, which is
satisfied in the $P\ll 1$ regime, i.e. when no more than a single
clump causes lensing of a background source: this results in
magnifications markedly larger than 1, or strong lensing. Smaller
magnifications, including de-magnifications, attain contributions
from the distribution of dark matter along the entire line of
sight, resulting in weak lensing effects. The latter case can
be represented by a Gaussian probability per unit magnification:
\begin{equation}
\label{weakpdiA}
  p(\mu,z)=H(z)\,\exp[-(\mu-\bar\mu)^2/2\sigma_\mu^2(z)]\; .
\end{equation}
where the location of the peak, $\bar\mu$, and the amplitude,
$H(z)$, are determined by the normalisation and flux conservation
conditions obtained by integrating over all possible
magnifications the combined (weak plus strong lensing) probability
distribution: $\int d\mu\, p(\mu ,z)=\int d\mu\, \mu p(\mu ,z)=1$.
The transition between the weak and strong lensing regimes is set at a
suitable magnification $\mu_{\rm cut}$: a convenient choice is
$\mu_{\rm cut}=1+1.5\sigma_\mu(z)$, yielding $\mu_{\rm
cut}\approx1.5-2$ for the redshift range of interest.

The maximum magnification for strong lensing, $\mu_{\rm max}$, determined by
the intrinsic  size of the source (Peacock 1982) and by the geometry of the
lens configuration was found to lie in the range 10-30 in the case of
our model (Perrotta et al. 2001). Here we adopt the conservative
value $\mu_{max}=10$.

Lens density profiles have been modelled as Singular Isothermal
Spheres (SIS) for masses $M < 3.5 \times 10^{13}\ M_{\odot }$, and
with the Navarro, Frenk \& White (1997; NFW) formula for larger
masses. In fact, it is found that (Porciani \& Madau 2000) such a
``mixed'' model provides a good fit to the  observed statistics of
the angular splitting of QSO multiple gravitational images.
However we wish to emphasise that regarding statistical
magnifications SIS and NFW profiles lead to comparable results
(Perrotta et al. 2001), making the present conclusions very weakly
dependent on the choice between SIS and NFW.

Before going to the results, let us consider the magnification bias on a
flux-limited source sample. The integrated source counts above a flux density
threshold $S_\nu$ of sources with a comoving luminosity function $\Phi(L,z)$
can be written as (e.g.~De Zotti et al.~1996)
\begin{equation}
\label{sourcecounts}
  N(S_\nu)=\int_0^{z_0}\,{\rm d}z\int_{L_{\rm min}}^\infty\,{\rm d}L\,
  \Phi(L,z)\,r^2(z)\,
  \frac{{\rm d}r}{{\rm d}z}\,\mbox{sr}^{-1}\;,
\end{equation}
where $r$ is the comoving radial distance, and
\begin{equation}
\label{Lmin}
  L_{\rm min}(\nu)=4\pi(1+z)\,r^2(z)\,S_\nu\,
  \frac{L(\nu)}{L[(1+z)\nu]}\;.
\end{equation}
The luminosity function modified by the magnification bias reads
(e.g.~Pei 1995):
\begin{equation}
\label{phieffective}
  \Phi'(L,z)=\int_{\mu_{\rm min}}^\infty{\rm d}\mu\,
  \frac{p(\mu,z)}{\mu}\,\Phi\left(\frac{L}{\mu},z\right)\;.
\end{equation}
Lensing effects on the source counts are taken into account by
replacing $\Phi'(L,z)$ with $\Phi(L,z)$ in Eq.~(\ref{sourcecounts}).

\section{Effect of gravitational lensing on source counts}
\label{results}

We present here our estimates for the effect of lensing on
integral number counts predicted using the model by Granato et al.
(2001) and, for comparison, using the model by Rowan-Robinson (2001),
taken as representative of the most successful phenomenological
models. The model by Granato et al. (2001) includes three populations, 
namely  spheroids,  slowly-evolving starburst and evolving disk galaxies 
(spirals); the model by Rowan-Robinson instead includes AGN, spirals, 
high-luminosity and low-luminosity starburst. In both models,  
we added the contribution due to ``flat''-spectrum radio
sources, whose counts (kindly provided by L. Toffolatti) are based
on the model by Toffolatti et al. (1998). 


The contributions of different populations to the number counts at
fluxes probed by SCUBA predicted by the two models are markedly
different: while, as discussed above, according to Granato et al.
(2001) SCUBA sources are mostly massive dusty spheroids at high
$z$, in the Rowan-Robinson (2001) model there is a large tail
extending to low-$z$. Correspondingly, while the model by Granato
et al. (2001) predicts an essentially exponential decline of
$850\,\mu$m counts with increasing flux density above a few mJy,
the slope is substantially less steep in Rowan-Robinson's (2001)
model.

This different behaviour of the counts explains the vastly
different fractions of lensed galaxies expected at bright fluxes,
as quantified in Figs.~\ref{cg} and \ref{crr}, which show the
integral number counts of lensed and unlensed galaxies, yielded by
the two models, respectively, for the $550$ (top), $850$ (middle)
and $1380\,\mu$m (bottom) {\sc Planck}/HFI channels. The heavy
solid lines correspond to strongly gravitationally lensed sources
(magnifications $\mu > 2$). Note that, for the model of Granato et al. 
(2001), forming spheroids give the relevant contribution to lensed 
counts, due to their exponential slope in number counts: thus, the solid 
lines in Fig. \ref{cg} are given by the lensed spheroids, 
while  starburst, spirals and radio sources are included only in the total 
counts (sum of weakly and strongly lensed sources), represented in  
Fig. \ref{cg} by the heavy dotted lines.
 
In the  case of Rowan-Robinson (2001) all classes of sources (AGN, 
spirals and starburst) are taken into account in the strongly-lensed 
counts of Fig. \ref{crr} (solid lines). The heavy dashed lines show the 
counts for these classes of sources, obtained by only allowing for weak lensing 
(magnifications $\mu < 2$ and de-magnifications), while the thin dotted 
lines (almost coincident with the heavy dashed lines in the case of
Rowan-Robinson 2001) show the counts computed if one ignores the
effect of lensing. Finally, the heavy dotted lines give the sum of
weakly and strongly lensed sources, also added with the contribution of 
radio  sources, for which the effect of lensing is negligible. 

It is clear from Fig.~\ref{cg} that, according to the model by
Granato et al. (2001) and thanks to the strong magnifications by
gravitational lensing, even faint forming spheroids at substantial
redshifts can be detected in relatively shallow surveys, such as
those provided by P{\sc lanck}/HFI. The quoted $5\sigma$ detection
limits for this instrument are $S_{\rm lim}=200,100,450$ mJy at
1380,850,550 $\mu$m, respectively. Note that, at 850 $\mu$m, this
implies that the expected strongly lensed objects at fluxes larger
than $S_{\rm lim}$ on the whole sky are $\sim 10^{2}$ in the
Rowan Robinson scenario (2001) and $\sim 4\cdot 10^{3}$ in the
model by Granato et al. (2001). Strongly lensed forming
spheroids may also be detectable by forthcoming balloon
experiments operating at sub-mm wavelengths, such as BLAST (Devlin
et al. 2000; www.hep.upenn.edu/blast/) and ELISA. Such surveys may
then trace the large scale distribution of the peaks of the
primordial density field.

Note that the distribution of strongly lensed sources would extend
to even higher fluxes in the case of magnifications larger than
our conservative limit $\mu_{max}=10$ (magnifications of up to 30
are possible according to the adopted model; cf. Perrotta et al.
2001).

Fig.~(\ref{ratio}) compares the fraction of lensed to unlensed
sources as a function of 850~$\mu$m flux predicted by the two
models considered above. According to the model by Granato et al.
(2001), at $850\,\mu$m the ratio of strongly lensed to unlensed
sources reaches the value of about 40\% for fluxes slightly below
$100\,$mJy, where the surface density of strongly lensed sources
is of about $0.1\,\hbox{deg}^{-2}$. On the other hand, the
fraction of strongly lensed sources is always below 10\%  for the
model of Rowan-Robinson (2001). It is then clear that
gravitational lensing can help to discriminate different
models, all of them successfully reproducing the observed counts.

In principle, clustered lenses can change the clustering properties of
the sources in a flux-limited sample. The auto-correlation function of
the lensing magnification pattern can be described as a convolution of
the magnification pattern of a single lens with the correlation
function of the lens centres. In our case, the magnification patterns
of individual lenses have typical angular scales of at most a few arc
seconds. The auto-correlation of lens centres, however, has a typical
angular scale of arc minutes, and its amplitude is substantially
lowered by projection because the lens population is taken from a
large redshift range. Therefore, the effect of lensing on the
clustering properties of sub-mm sources is negligible.

\section{Conclusions}
A major challenge for current theories of galaxy formation is to
account for the evolutionary history of large spheroidal galaxies.
Even the best semi-analytic models (Cole et al. 2000; Devriendt \&
Guiderdoni 2000) hinging upon the ``standard'' picture for
structure formation in the framework of the hierarchical
clustering paradigm, tuned to agree with detailed numerical
simulations, are stubbornly unable to account for the sub-mm
counts of galaxies as well for the (somewhat still controversial)
optical evidence for most large ellipticals we see today were
already in place at $z\simeq 2$ and (almost) passively evolving
since then (Daddi et al. 2000; Cohen 2001). These data are more
consistent with the traditional approach whereby elliptical
galaxies form most of their stars in a single gigantic star-burst
followed by essentially passive evolution (``monolithic'' models).
This approach, however, is inadequate to the extent that it cannot
be fitted in a consistent scenario for structure formation from
primordial density fluctuations.

A possible way out was proposed by Granato et al. (2001) who
worked out a physically motivated scheme according to which the
mm/sub-mm galaxies detected by SCUBA and MAMBO surveys are large
proto-ellipticals in the process of forming most of their stars
and of growing a quasar at their centres. In their model, the
early evolution of these galaxies and of quasars are tightly
inter-related and feed-back effects (mostly due to supernova
explosions) make large ellipticals to form as soon as the
corresponding potential wells are in place, while the formation of
smaller galaxies is delayed. Therefore, in this model, large
ellipticals evolve essentially as predicted by ``monolithic''
models and the mm/sub-mm counts are successfully reproduced. The
star-formation activity, powering the dust emission, quickly
declines for $z< 3$ for the most luminous (most massive) galaxies,
while quasars reach their maximum luminosity. This naturally
explains why very luminous quasars are more easily detected at
mm/sub-mm wavelengths at redshifts larger than that ($\simeq 2.5$)
of maximum quasar activity (Omont et al. 2001). As indicated by
the analysis of the latter authors, a substantial fraction of the
observed far-IR emission is probably powered by the starburst in
the host galaxy.

In this paper predictions of the model are extensively
investigated and compared with recent data. Attention is focused,
in particular, on two specific aspects. Since SCUBA/MAMBO galaxies
are interpreted as massive galaxies at $2\simlt z \simlt 6$, they
are expected to be highly biased tracers of the matter
distribution and therefore highly clustered. The implied angular
correlation function is found to be consistent with recent results
by Scott et al. (2001), Peacock et al. (2000), Lagache \& Puget
(2000) as well as with the fluctuations in the SCUBA counts in
different areas of the sky. Explicit estimates are presented for
the power spectrum of temperature fluctuations due to clustering
in {\sc Planck}/HFI channels.

A second specific prediction of the model is an essentially
exponential decline of the counts at $S_{850\mu{\rm m}} \simgt
10\,$mJy. Hints in this direction that can possibly be discerned
in recent SCUBA (Scott et al. 2001; Borys et al. 2001) and MAMBO
(Carilli et al. 2000) surveys are noted. As an implication of this
prediction, together with the fact that sources are found at high
redshifts, one has that both the gravitational lensing probability
and the magnification bias on the counts are much higher than
those derived for other current, phenomenological, models. We show
that, according to this model, essentially all proto-spheroidal
galaxies brighter than $S_{850\mu{\rm m}} \simgt 60$--70$\,$mJy
are gravitationally lensed. Allowing for the other populations of
sources contributing to the bright mm/sub-mm counts, we find that
the fraction of gravitationally lensed sources may be $\simeq
40\%$ at fluxes slightly below $S_{850\mu{\rm m}} = 100\,$mJy. If
so, large area surveys such as those to be carried out by {\sc
Planck}/HFI or by forthcoming balloon experiment like BLAST and
ELISA will probe the large scale distribution of the peaks of the
primordial density field.

\bigskip \noindent {\bf Acknowledgements}
\noindent We are grateful to P. Panuzzo and L. Toffolatti  for
help with the counts of spheroidal galaxies and radio sources,
respectively. Work supported in part by ASI, MURST and UE.

\end{document}